\documentclass[aps,prd,twocolumn,showpacs,preprintnumbers,a4paper]{revtex4}
\usepackage{dcolumn}
\usepackage{bm}
\usepackage{amssymb,amsmath}
\usepackage[english]{babel}
\usepackage[dvips]{graphicx}
\usepackage{subfigure}
\usepackage{color}
\usepackage{psfrag}

\def\lya{Ly$\alpha\ $}

\newcommand{\om}{\Omega_{\rm 0m}}

\newcommand{\sig}{\sigma_8}
\newcommand{\ns}{n_{\mathrm{s}}}
\newcommand{\nrun}{n_{\mathrm{run}}}

\newcommand{\gad} {{\small {GADGET-2}}\,}

\newcommand{\simgt}{\,\hbox{\lower0.6ex\hbox{$\sim$}\llap{\raise0.6ex\hbox{$>$}}}\,}
\newcommand{\simlt}{\,\hbox{\lower0.6ex\hbox{$\sim$}\llap{\raise0.6ex\hbox{$<$}}}\,}

\begin{document}
\thispagestyle{empty}

\preprint{LAPTH-1182/07, arXiv:0705.0533} 

\title{A combined analysis of 3D Weak Lensing, Lyman-$\alpha$ forest
and WMAP year three data}

\author{J. Lesgourgues $^1$\footnote{\rm{email: lesgourg@lapp.in2p3.fr}},  
M. Viel$^{2,3}$\footnote{\rm{email: viel@oats.inaf.it}}, 
M.G. Haehnelt $^4$, R. Massey $^5$}
\affiliation{$^1$ LAPTH, Universit\'e de Savoie \& CNRS (UMR5108), BP110,
F-74941 Annecy-le-vieux Cedex, France}
  \affiliation{$^2$
        INAF-Osservatorio Astronomico di Trieste, Via G.B. Tiepolo 11,
        I-34131 Trieste, Italy}
 \affiliation{$^3$ INFN/National Institute
        for Nuclear Physics, Via Valerio 2, I-34127 Trieste, Italy}
 \affiliation{$^4$
         Institute of Astronomy, Madingley Road, Cambridge CB30HA, England}
 \affiliation{$^5$
         California Institute of Technology, 1200 East California Boulevard, Pasadena, CA 91125, U.S.A.}

\date{\today}

\begin{abstract}
\noindent

We present constraints on the amplitude and shape of the matter power
spectrum and the density of dark matter within the framework of a
standard $\Lambda$CDM model. We use a Markov Chain Monte Carlo
approach to combine independent measurements of the three dimensional
weak gravitational lensing shear field by the COSMOS survey, of low
and high resolution \lya forest flux power spectrum by SDSS and LUQAS,
and of Cosmic Microwave Background temperature and polarization
anisotropies by WMAP. We note good agreement between the amplitude of
the matter power spectrum on intermediate and small scales as inferred
from \lya forest and lensing data. The \lya forest data helps to break
the $\sig-\om$ degeneracy characteristic of weak lensing results,
yielding $\sig = 0.876 \pm 0.048$ for COSMOS plus \lya SDSS data. This
is somewhat larger than the value preferred by the WMAP year three CMB
data. Combining all three data sets significantly tightens the
constraints on $\sig$, the spectral index of primordial density
fluctuation $\ns$, a possible running of the spectral index $\nrun$
and the matter density $\om$. Assuming no running, the joint
constraints for COSMOS, SDSS and WMAP are $\sig=0.800\pm0.023$,
$\ns=0.971\pm0.011$, $\om=0.247\pm0.016$ (1-$\sigma$ error bars).

\end{abstract}



\maketitle

\numberwithin{equation}{section}


\section{Introduction}
\label{intro}
The recent measurement of the Cosmic Microwave Background (CMB)
anisotropies obtained by the WMAP satellite
\cite{Spergel:2006hy,Page:2006hz,Hinshaw:2006ia} has considerably
tightened  the error bars on the cosmological parameters that 
describe the standard $\Lambda$CDM model of structure formation.  The
constraining power of the WMAP year three measurements alone is 
already large. Cosmological parameter extraction
nevertheless benefits from a  combination of the CMB data  with
measurements of the matter power spectrum on smaller scales and
at different redshifts. The various observables  suffer from very different
systematic and statistical errors, and a combined analysis can
break degeneracies inherent to individual measurements.

We concentrate here on comparing and combining the CMB data with weak
gravitational lensing and \lya forest data.  The \lya forest due to
the absorption produced by neutral hydrogen along the line-of-sight to
distant quasars (QSOs) allows us to measure the matter power spectrum
on scales that range from few to tens of comoving Mpc at $z=2-4$
(e.g. \citep{Croft:2000hs}). ``Cosmic shear'' measurements of the
distortion induced in distant galaxy images by weak gravitational
lensing around foreground structures map the mass distribution at
similar scales but smaller redshift, $z=0-1.5$. Both methods probe the
matter power spectrum on smaller scales than CMB anisotropies observed
by WMAP, and offer a more direct measurement of the quantity
$\sigma_8$, the r.m.s. of the density fluctuations in spheres of radii
8 comoving $h^{-1}$ Mpc. Furthermore, the small scale matter power
spectrum inferred from the \lya forest data puts strong limits on the
mass of warm dark matter candidates and on isocurvature perturbations
(e.g. \citep{Viel:2005qj,seljakmakarov,vielsterile,
seljak,Beltran:2005gr}).

Viel et al. \cite{viel06} and Seljak et al. \cite{seljak} have
recently presented independent analyses of CMB data combined with \lya
forest data from separate, state-of-the-art QSO samples: the LUQAS
sample of high-resolution high quality VLT-UVES spectra
(\citep{bergeron,Kim:2003qt,Viel:2004bf}) and a large sample of SDSS
QSO spectra (\citet{McDonald:2004eu}).  Despite the very different
data sets and the use of different analysis techniques, both groups
found similar results, suggesting a value for $\sigma_8\sim0.9$ larger
than that extrapolated from the CMB data alone
($\sigma_8=0.76\pm0.05$).  For the case of the SDSS \lya forest data
there appears to be a moderate tension between the two data sets at
the 2 $\sigma$ level~\cite{seljak}.  For the LUQAS data the
errors are about a factor two larger and the difference is not
statistically significant~\cite{viel06}. Ref.~\citep{Jena:2004fc}
obtained very similar results with a further \lya forest data set.

The moderate but notable tension between the amplitude of the matter
power spectrum inferred from the SDSS \lya forest data with the WMAP
three year results suggests that measurements of the amplitude of the
small and intermediate scale matter power spectrum are still somewhat
uncertain. Further comparison with a third, completely independent
technique should be very useful in this context. We exploit the
tomographic weak gravitational lensing analysis used to map the three
dimensional distribution of mass in the {\it Hubbble Space Telescope}
COSMOS survey \citep{massey,massey2}. The clustering signal in that
distribution also corresponds to a relatively large value of
$\sig=0.95^{+0.093}_{-0.075}$ (when the matter fraction $\om$ is
chosen to be the WMAP maximum likelihood value). A recent combination
of two dimensional, ground-based cosmic shear surveys into the 100
square degree Weak Lensing Survey \citep{benjamin} gives a comparable
although slightly smaller amplitude of the matter power spectrum
$\sig=0.84\pm0.07$ (for the best fit $\om$ WMAP value).

In this paper, we investigate in detail to what extent the \lya forest
and the cosmic shear data of Massey et al. (2007) are consistent, and
present new constraints on the value of cosmological parameters
obtained from various combinations of cosmic shear, \lya forest and
CMB data (for the latter we limit ourselves to the WMAP year three
determination of temperature/polarization anisotropies
\cite{Spergel:2006hy}). We use the Boltzmann code {\sc
camb}~\cite{Lewis:1999bs} for computing the linear matter power
spectrum and applying relevant non linear corrections \cite{smith}.
The multi dimensional parameter space is explored with Monte Carlo
Markov Chains (MCMC) using the public code {\sc cosmomc}
\citep{lewis}.


\section{The data sets}
\label{data}
\subsection{\lya forest data}

We have used two \lya forest data sets: {\it i)} the high resolution
QSO absorption spectra presented in 
Viel, Haehnelt \& Springel (VHS) \cite{Viel:2004bf} and in
Refs.~\citep{Viel:2004np,viel06}, consisting of the LUQAS sample
\citep{Kim:2003qt} and the reanalyzed Croft et al. (C02)
\cite{Croft:2000hs} data; {\it ii)} the SDSS \lya forest sample
presented in \citet{McDonald:2004eu} (M05).  The SDSS \lya forest data
set consists of $3035$ QSO spectra with low resolution ($R\sim 2000$)
and low S/N ($< 10$ per pixel) spanning a wide range of redshifts
($z=2.2-4.2$), while the LUQAS and the C02 samples contain mainly 57
high resolution ($R \sim 45000$), high signal-to-noise ($>50$ per
pixel) QSO spectra with median redshifts of $z=2.125$ and $z=2.72$,
respectively.  

The flux power spectrum of the \lya forest is the quantity which is
observed and needs to be modeled at the percent or sub-percent level
using accurate numerical simulations that incorporate the relevant
cosmological and astrophysical processes. M05 modeled the flux power
spectrum using a large number of Hydro Particle Mesh (HPM)
simulations~\citep{Gnedin:1997td,Viel:2005eg}, calibrated with full
hydrodynamical simulations. Instead, the VHS analysis significantly
improved the effective bias method developed by C02
(see~\citet{Gnedin:2001wg} and \citet{Zaldarriaga:2001xs} for a
critical assessment of the errors involved), by using a grid of full
hydrodynamical simulations run with the Tree-SPH code \gad
\citep{Springel:2000yr, Springel:2005mi} to infer the linear matter
power spectrum. Finally, \citet{Viel:2005ha} performed an independent
analysis of the SDSS \lya forest data, and used a Taylor expansion of
the flux power spectrum around best fitting values based on full
hydrodynamical simulations to model the dependence of the flux power
on cosmological and astrophysical parameters.  This analysis was
performed directly on the flux power spectrum and took thus full
advantage of each data points, for a wider range of parameters than
just the amplitude and slope constrained by the SDSS analysis. One
should however keep in mind that the Taylor expansion approach is
likely to underestimate the errors far from best-fit values.
However, as soon as the SDSS Lyman-$\alpha$ data is combined with either
COSMOS or WMAP, large departures from the best-fit model are
forbidden and the Taylor method is accurate.

In this paper, we will use either the data set of VHS, based on
high-resolution QSO spectra and noted ``LyaVHS''; or the results of
\citet{Viel:2005ha} using low-resolution SDSS spectra and a Taylor
expansion, noted ``LyaSDSS-d'' (where -d refers to ``derivatives'',
since this method is based on the derivatives of the flux
power spectrum.

Our {\sc cosmomc} module {\tt lya.f90} comparing the linear dark
matter power with LyaVHS data has been incorporated into the latest
public available version of {\sc cosmomc} \citep{lewis}.  The LyaVHS
power spectrum consists of estimates of the linear dark matter power
spectrum at nine values in the wavenumber space $k$ at $z=2.125$ and
nine values at $z=2.72$, in the range $0.003<k$ (s/km)$<0.03$.  The
estimate of the uncertainty of the overall amplitude of the matter
power spectrum is $29\%$. This estimate takes into account possible
systematic and statistical errors (see the relevant tables of VHS for
a detailed discussion). The code assigns a Gaussian prior to the
corresponding nuisance parameter and marginalize over it.  For the
LyaSDSS-d analysis, we used the {\sc cosmomc} module described in
\cite{viel06}, which involves 21 nuisance parameters characterizing a
wide range of astrophysical and noise-related systematic
uncertainties.  In the following results these parameters are always
marginalized out.

\subsection{Weak lensing data}

The {\it Hubbble Space Telescope} COSMOS survey covers a contiguous
area of 1.64 square degrees on the sky. In this high resolution,
space-based data, the shapes of 234,370 distant galaxies were
measured, with a median $F814W_{AB}$-band magnitude of 24.6 
\citep{apjse_lea,massey}.

A crucial addition to this survey has been the acquisition of
ground-based imaging in 15 extra bands. Photometric redshift
estimation for each galaxy allows a fully 3D exploitation of the
signal, in which the power spectrum can be independently measured at
different redshifts and physical scales. In a 2D analysis, these would
have been projected together, resulting in a loss of discriminatory
power on cosmological parameters of a factor of $3-5$
\citep{massey,heavens}. BPZ photometric redshift estimation software
achieved $68\%$ confidence limits of $0.03(1+z)$ on each galaxy to
$z\sim 1.4$ and $I_{F814W}=24$ \citep{apjse_mob}. The galaxy catalog
has a median photometric redshift of $z_{\rm phot}=1.26$, and has been split into
three redshift bins for this analysis: $z_{\rm phot}=0.1-1$, $z_{\rm phot}=1-1.4$ and
$z_{\rm phot}=1.4-3$, which divide the number of sources almost evenly.

In each redshift bin, the ``cosmic shear'' two-point correlation
functions $C_{1,2}(\theta)$ have been measured on angular scales
$0.1-40$ arcmin \citep{massey}. Error estimates for these measurements
include: statistical errors, calculated from the internal distribution
of shear estimators; cosmic (sample) variance, calculated from the
variation in the signal between separate quadrants of the COSMOS
field; and finally systematic errors, which include the potential bias
in shear calibration (overall bias and relative bias between redshift
bins), errors due to catastrophic photometric redshift failures, and
errors due to binning.

In Massey et al. \cite{massey}, the constraints on cosmological
parameters were derived as follows. For a three-dimensional grid of
models spanning variations of $\om$ from 0.05 to 1.1, $\sigma_8$ from 0.35 to
1.4 and the power spectrum shape parameter $\Gamma$ from 0.13 to 0.33,
the linear power spectra were obtained from the fitting formula of
BBKS \cite{BBKS}, and corrected for non-linear evolution using {\sc
halofit} \cite{smith}. For each model, the data likelihood was
computed taking only statistical errors into account. Then, the
three-dimensional likelihood distribution was integrated in order to
marginalize over $\Gamma$ and to obtain confidence contours for
($\om$, $\sigma_8$). The best constrained combination of these
parameters was found to be $\sigma_8 (\om/0.3)^{0.44}$. Final bounds
on this quantity were obtained by adding systematic error linearly.

In this work, we wrote a {\sc cosmomc} module for COSMOS data
(downloadable at {\tt http://www.astro.caltech.edu/\~{ }rjm/cosmos/cosmomc/})
which offers various advantages with respect to the original analysis:
the linear power spectrum is computed by {\sc camb}, more free
parameter can easily be implemented in the analysis (like the tilt
$n_S$), and the systematic errors can be accounted more accurately by
introducing various nuisance parameters. The module computes the shear
correlation functions for any cosmological model explored by the
Markov chains (including corrections to the linear power spectrum
obtained with the {\sc halofit} code \footnote{ The authors of
Ref.~\cite{massey} checked that using Peacock and Dodds fitting
functions instead of {\sc halofit} only affects the result on
$\sigma_8$ by $\sim$5\%. Note also that on these scales, one might
investigate the impact of baryon physics on the matter power spectrum
\cite{Rudd:2007zx}, although this effect is still uncertain.})  and
compares with the tomographic results from Ref.~\citep{massey},
neglecting the cross-correlation between redshift slices. As shown in
figure 8 of Ref.~\citep{massey}, there are measurements on six angular
scales, in each of three redshift bins. Since there are two
correlation functions available for shear, the number of data points
sums up to 36. The $36\times 36$ covariance matrix is presented in
figure 9 of \citep{massey}, and we used its inverse when calculating
the likelihood.  We modeled the systematics errors of the lensing data
described in detail in Ref.~\cite{massey} by three nuisance
parameters, over which our final results are marginalized. A blind
analysis of simulated HST images suggests a potential 6\% uncertainty
in the overall calibration of the shear measurement \cite{apjse_lea},
for which we account with a parameter $A$ with Gaussian prior.  We
further introduce a parameter $B$ to account for a 5\% relative
calibration uncertainty between the shear measured from galaxies in
our high and low redshift bins.  Although not seen in the simulated
HST simulations of \cite{apjse_lea}, more comprehensive tests on a
larger set of simulated ground-based images \cite{step1,step2} reveal
the potential for the shear to be underestimated in faint or small
galaxies. The opposite effect has not been recorded. $B$ is thus
assigned a one-sided Gaussian prior.  Finally, a potential 10\%
intrusion of low redshift galaxies into the high redshift bin due to
catastrophic photometric redshift errors is modeled with a third
nuisance parameter $C$. Priors on the photometric redshifts were
designed to ensure that the effect on lensing observables of known
types of catastrophic failure are easily modeled. The failures can
only dilute the signal in the high redshift bin, and again we use a
one-sided Gaussian prior. In summary, the data points
$C_{1,2}(\theta)$ are multiplied by
\begin{eqnarray}
& (A/B)^2   & \qquad (\mathrm{low-}z) \nonumber \\
& A^2      & \qquad (\mathrm{medium-}z) \nonumber \\
& (AB)^2/C & \qquad (\mathrm{high-}z) \nonumber
\end{eqnarray}
with priors on $A$, $B$ and $C$ peaking at  one with 
$B \geq 1$, $C \geq 1$, $\sigma_A=0.06$, $\sigma_B=0.05$ and
$\sigma_C=0.10$.

\section{Results}
\label{results}

\begin{figure*}
\includegraphics[angle=0,width=12cm]{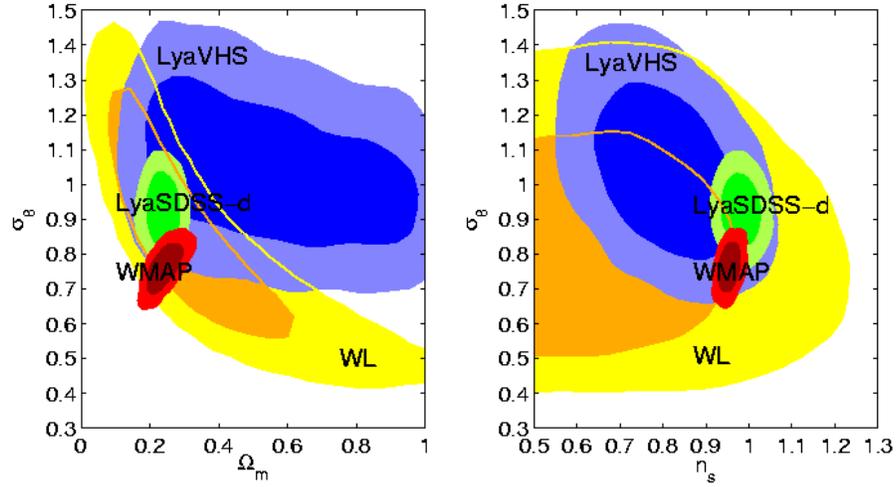}
  \caption{\label{fig_each} 1-$\sigma$ and 2-$\sigma$ 
   contours of the marginalized likelihood
   in the $\sig-\om$ plane (left) and $\sig-\ns$ plane (right)
  for each data set separately: WMAP (red), weak lensing
  (orange), \lya forest from VHS~\cite{Viel:2004bf} (blue) and from
  SDSS as analyzed by \cite{Viel:2005ha} (green).}.
\end{figure*}

\begin{figure*}
\includegraphics[angle=0,width=12cm]{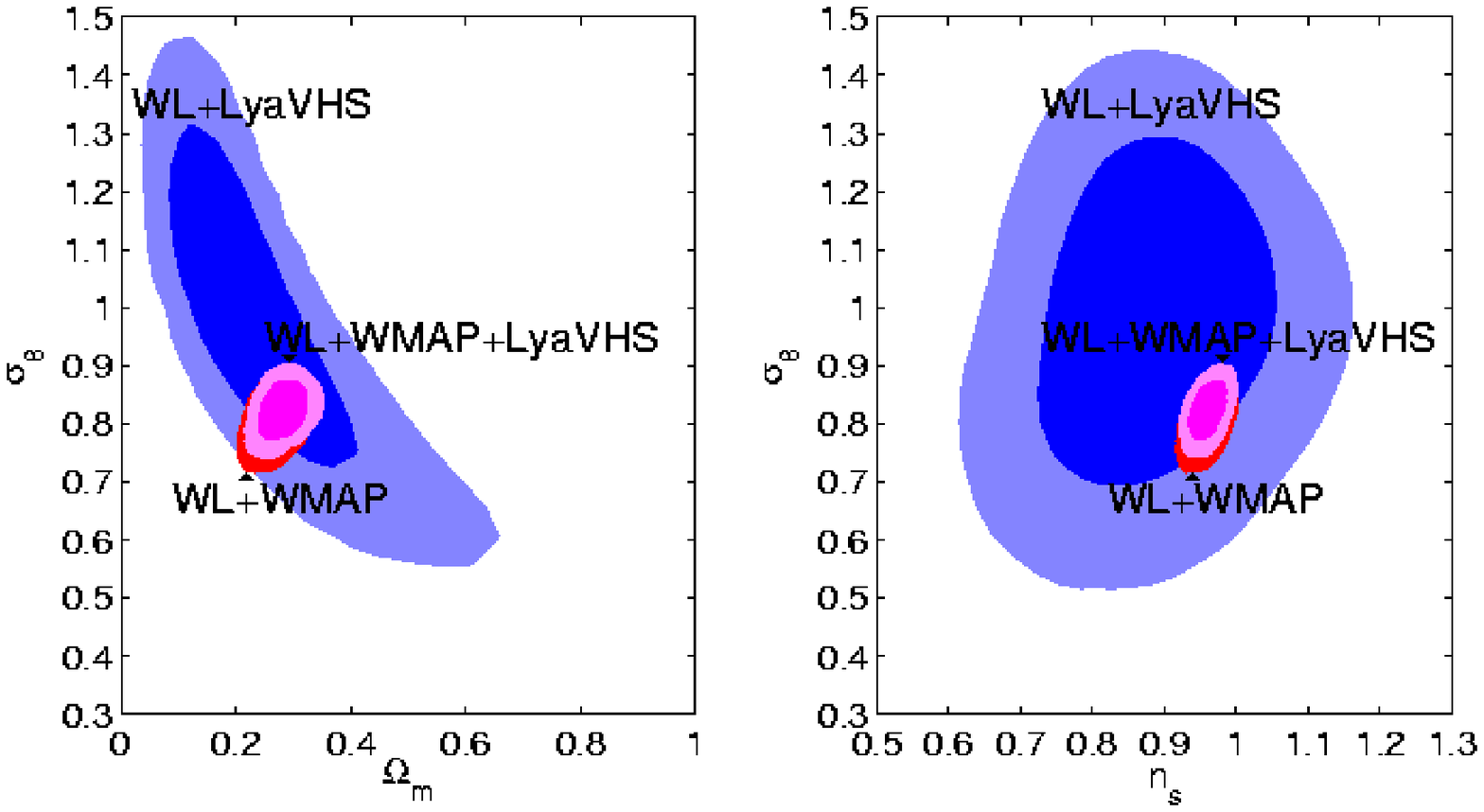}\\
\includegraphics[angle=0,width=12cm]{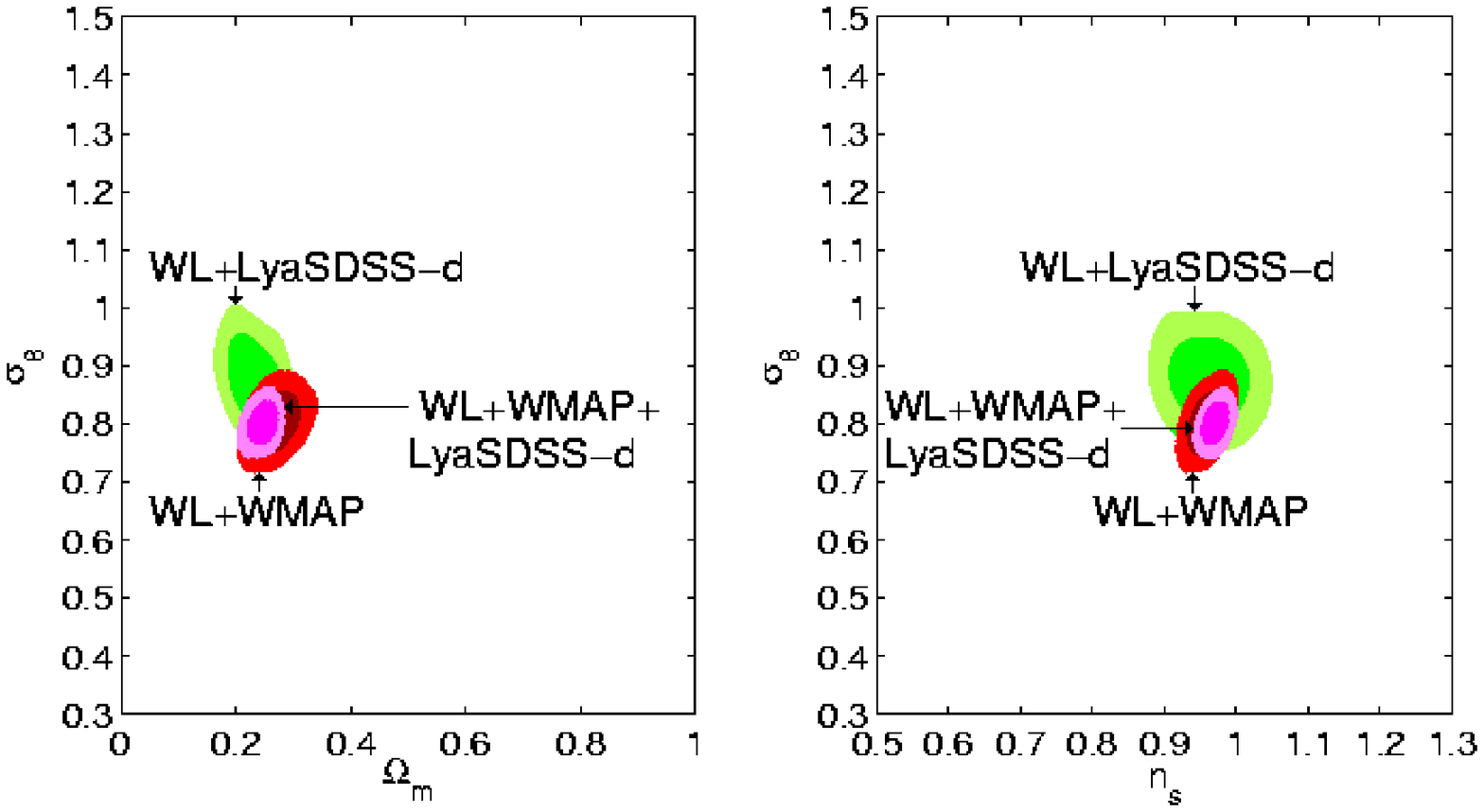}
  \caption{\label{fig_comb} 1-$\sigma$ and 2-$\sigma$ contours of the marginalized
  likelihood  in the $\sig-\om$ plane (left column) and $\sig-\ns$ plane
  (right column) for various combinations of data sets from COSMOS (WL),
  WMAP, and \lya forest data from VHS~\cite{Viel:2004bf} (LyaVHS, upper
  panels) and SDSS as analyzed by \cite{Viel:2005ha} (LyaSDSS-d, lower
  panels).}
\end{figure*}

For our MCMC analysis we have assumed a minimal flat $\Lambda$CDM
model, with no tensor contribution.  We vary the following
cosmological parameters with top-hat priors: dark matter
density $\Omega_{0c} h^2 \in [0.01,0.99]$, baryon density $\Omega_{0b} h^2
\in [0.005,0.1]$, primordial spectral index $n_s \in [0.5,1.5]$,
primordial amplitude $\log[10^{10} A_s] \in [2.7,4.0]$ and angular
diameter of the sound horizon at last scattering $\theta \in
[0.5,10]$. When CMB data is included, we also vary
the optical depth
to reionization $\tau \in [0.01,0.8]$. For part of
the MCMC analysis, we have finally considered a running of the spectral
index, $\nrun \in [-0.5,0.5]$. From the parameters above the MCMC
code derives the reduced Hubble parameter $h$, the matter fraction
$\om$ and $\sig$: so, these parameters have non-flat
priors and the corresponding bounds must  be interpreted with some care. In
addition, {\sc cosmomc} imposes a weak prior on the Hubble parameter: $h \in
[0.4,1.0]$. In each case and for the purpose of comparing with
Ref.~\cite{massey}, we also compute the combination $\sigma_8
(\om/0.3)^{0.44}$, which is best probed by the COSMOS lensing
data.

We ran first a MCMC analysis of the COSMOS WL data alone. We recall
that this analysis differs from that in Massey et al. \cite{massey} in
four ways: the methodology (MCMC with flat priors on the cosmological
parameters instead of maximum likelihood), the introduction of one
extra independent parameter $\ns$, the fact that we compute the exact
linear matter power spectra numerically, and the treatment of
systematic errors. In Ref.~\cite{massey}, the analysis performed
with statistical errors only led to the bounds $\sig
(\om/0.3)^{0.44}=0.866\pm0.033$ (68\% Confidence Level (C.L.)), which
extend to $\sigma_8 (\om/0.3)^{0.44}=0.866^{+0.085}_{-0.068}$ when
systematic errors are added linearly.  Here we first performed a MCMC
analysis excluding the systematic errors, and obtained $\sig
(\om/0.3)^{0.44}=0.83\pm0.06$. The increased error bar with respect to
\cite{massey} is presumably due to the inclusion of an arbitrary
spectral index $\ns$ in the analysis, which opens new parameter
degeneracies.  The mean value of Massey et al. (2007) (0.866) is
perfectly consistent with this value.  We then
incorporated systematic errors as described in the last section and
found
\begin{equation}
\sigma_8 (\om/0.3)^{0.44} = 0.814 \pm 0.074 \qquad (68\%{\rm C.L.}) ~.
\end{equation}
Note that because
of non-linear corrections to the matter power spectrum, the $\sigma_8$
parameter cannot be viewed as a simple calibration parameter for the
theoretical correlation functions. A change in $\sigma_8$ changes both
the amplitude and the shape of the shear correlation functions in a
non-trivial way. As a consequence, the impact of systematic errors on
the determination of the (linear theory) parameter $\sigma_8
(\om/0.3)^{0.44}$ is found to be smaller than the data calibration
uncertainty itself. The 68\% confidence limits on each parameter are
presented in the first column of Table~I.  We also show the joint 68\% and 95\%
confidence contours in $\sig$--$\om$ and $\sig$--$\ns$ space
in Fig.~\ref{fig_each} (yellow). 

\begin{table}
\begin{center}
\caption{Summary of the constraints on $\sig$, $\ns$, $\om$, $h$ and
$\tau$, for the minimal 6-parameter $\Lambda$CDM model and each data
sets. Since this is a Bayesian analysis, the bounds depend on our choice of
priors; our top-hat priors are described at the beginning of Sec.~\ref{results} (in
particular, we impose a weak $h$ prior: $0.4<h<1.0$). The quoted values are either 
the mean and 68\%~C.L. error,
or only the 68\%~C.L. upper/lower limit when a parameter is not bounded on both
sides within the prior range.}
\begin{tabular}{cccccccccc}
\hline
&  \tiny{WL} & \tiny{\lya VHS}  & \tiny{\lya SDSS-d} & \tiny{WMAP3} \\
\hline
$\sig$ 
& $~0.85\pm0.22$ & $~1.04\pm0.16$ & $~0.926\pm0.066$  & $~0.762\pm0.046$ \\
$\ns$  
& $~<0.94$       & $~0.80\pm0.10$ & $~0.982\pm0.028$  & $~0.955\pm0.016$ \\
$\om$  
& $~0.34\pm0.19$ & $~0.55\pm0.26$ & $~0.238\pm0.030$  & $~0.243\pm0.032$ \\
$h$    
& $~>0.71$       & $~>0.63$       & $~0.710\pm0.071$  & $~0.729\pm0.030$ \\
$\tau$    
& -- & -- & -- & $~<0.104$ \\
\hline
\label{tab1}
\end{tabular}
\end{center}
\end{table}

\begin{table}
\begin{center}
\caption{Same as table \ref{tab1} for the combination of weak lensing data
with each other data set.}
\begin{tabular}{cccccccccc}
\hline
& \tiny{WL+WMAP3} & \tiny{WL+\lya VHS} & \tiny{WL+\lya SDSS-d} \\
\hline
$\sig$ 
& $~0.802\pm0.034$ & $~0.98\pm0.19$ & $~0.876\pm0.048$\\
$\ns$  
& $~0.958\pm0.016$ & $~0.88\pm0.11$ & $~0.962\pm0.034$ \\
$\om$  
& $~0.269\pm0.026$ & $~0.25\pm0.12$ & $~0.232\pm0.028$\\
$h$    
& $~0.708\pm0.023$ & $~>0.79$       & $~0.773\pm0.089$\\
$\tau$    
& $~<0.103$ & -- & -- \\
\hline
\label{tab1bis}
\end{tabular}
\end{center}
\end{table}

\begin{table}
\label{tab1ter}
\begin{center}\caption{Same as table \ref{tab1} for the combination of
CMB, weak
lensing and \lya forest data. The
quoted values  are the mean and 68\% confidence limits.}
\begin{tabular}{cccc}
\hline
& \tiny{WL+WMAP3+\lya VHS}  &\tiny{WL+WMAP3+\lya SDSS-d} \\
\hline
$\sig$  & $~0.822\pm0.032$  & $~0.800\pm0.023$ \\
$\ns$   & $~0.960\pm0.016$  & $~0.971\pm0.011$ \\
$\om$   & $~0.282\pm0.026$  & $~0.247\pm0.016$ \\
$h$     & $~0.700\pm0.022$  & $~0.730\pm0.016$ \\
$\tau$  & $~0.094\pm0.028$  & $~0.109\pm0.026$ \\
\hline
\end{tabular}
\end{center}
\end{table}

\begin{table}
\label{tab2}
\begin{center}\caption{Same as table \ref{tab1ter}
for the case with a running 
spectral index.  The quoted values  are the mean and 68\% confidence limits.}
\begin{tabular}{ccccccccc}
\hline
& \tiny{WL+WMAP3+\lya VHS} & \tiny{WL+WMAP3+\lya SDSS-d} \\
\hline
$\sig$  & $~0.809\pm0.041$ & $~0.818\pm0.024$ \\
$\ns$   & $~0.965\pm0.018$ & $~0.971\pm0.015$ \\
$\om$   & $~0.304\pm0.032$ & $~0.255\pm0.018$ \\
$h$     & $~0.679\pm0.026$ & $~0.719\pm0.018$ \\
$\tau$  & $~0.085\pm0.037$ & $~0.135\pm0.026$ \\
$\nrun$ & $~-0.028\pm0.018$ & $~-0.007\pm0.021$ \\
\hline
\end{tabular}
\end{center}
\end{table}

For comparison, we also show in Fig.~\ref{fig_each} the $\sig$-$\om$ and
$\sig$-$\ns$ confidence regions for each dataset separately. The COSMOS
data is compatible with WMAP, since the contours have some overlap even at
the 68\% level. In the $\sig$-$\om$ space, the COSMOS and WMAP contours
appear as almost orthogonal, and the overlap clearly suggests that the
WL data prefers the highest $\sig$ values allowed by WMAP. The COSMOS
data is also compatible with the different \lya data sets, with again
an overlap at the 68\% level. Finally, the WMAP and \lya contours
overlap only at the 2 $\sigma$ level, as expected from previous
works~\cite{viel06,seljak}.

The two panels in Fig.~\ref{fig_each} provide a good illustration of
the advantages of  combining various datasets.  For each type of
experiment, $\sig$ and $\om$ are clearly correlated, but the direction
of the degeneracy is different for CMB, WL and \lya data. There are
various reasons for the difference between the direction of
correlation associated with the WL and \lya data: first, the raw \lya
data are in units of s/km (since the power spectrum is measured in
velocity space), and the rescaling to units of $h$/Mpc depends on
$\om$; second, the data probe  different redshifts, and the ratio
between the power spectrum today (when $\sig$ is defined) and at a
given redshift depends on $\om$; third, the slope of the matter power
spectrum depends on $\om$. This explains the ``banana shapes'' in the
upper left panels, with different orientations.

We don't find any significant correlation between $\sig$ and $n_s$ (upper
right panel in Fig.~\ref{fig_each}). For the WL and \lya data, this is due
to the fact that these experiments directly probe power on the 
scale at which $\sig$ is defined (if this was not the case, the
amplitude of the WL and \lya experimental points would constrain a
combination of $\sig$ and $\ns$).

We then ran some Markov chains for various combinations of different
data sets and we show the results in Table~\ref{tab1bis}, \ref{tab1ter} and
Fig.~\ref{fig_comb}. 
Combining the data sets significantly tightens the
constraints. Most noteworthy is that there remains a (moderate) strain
between the inferred value of $\sig$ between that inferred from WMAP
alone and that inferred from the lensing and \lya forest data. 
The constraint on $\sig$ from
COSMOS+Ly$\alpha$VHS ($\sig = 0.98 \pm 0.19$) and from
COSMOS+Ly$\alpha$SDSS-d ($\sig = 0.876 \pm 0.048$) are compatible with
the WMAP best-fit value ($\sig = 0.762$) respectively at
the 1.1-$\sigma$ and 2.4-$\sigma$ level.
Note that the SDSS-d
contours  are based on extrapolations using a Taylor
expansion of the flux power spectrum around a best fit model and are
likely to underestimate the error for parameters far from the
best fit model. However, when this data set is
used in combination with WL and/or WMAP data, the 68\%
and 95\% C.L. contours remain in a small region where the Taylor expansion
is accurate.
The main results of this work are the 68\% confidence limits for the
combined analysis of CMB, weak lensing and \lya forest data:
WMAP+COSMOS+Ly$\alpha$VHS, $\sig=0.822\pm0.032$;
WMAP+COSMOS+Ly$\alpha$SDSS-d, $\sig=0.800\pm0.023$. 

Finally, we performed a further MCMC analysis with an extended
$\Lambda$CDM model with on extra parameter, a running of the spectral
index $\nrun$.  The results are summarized in Table~IV. In this case,
the tilt $\ns$ is defined at the pivot scale $k_0=0.01~{\rm Mpc}^{-1}$
(when WL and \lya data are included, the pivot value $k_0=0.002~{\rm
Mpc}^{-1}$ adopted in the WMAP3 paper \cite{Spergel:2006hy} is too
small with respect to the median scale of the full data set). The
choice of a given pivot scale is indifferent for the definition of
$\nrun$.  WMAP alone is compatible with a rather large negative
running, $\nrun=-0.055\pm0.03$ (68\%C.L.) which results in a reduction
of power on small scale. Small scale experiments like \lya forest and
weak lensing observations are obviously crucial for the determination
of a possible running of the spectral index, since they increase the
lever arm for primordial spectrum reconstruction. The high $\sig$
value preferred by WL and \lya data excludes the most negative values
of $\nrun$ found by WMAP, and reduce its error by a factor of two. We
find $\nrun = -0.028\pm0.018$ for WMAP+COSMOS+Ly$\alpha$VHS and 
$\nrun = -0.007\pm0.021 $ for WMAP+COSMOS+Ly$\alpha$SDSS-d (68\%C.L.).


\section{Conclusions}
\label{conclusions}
We have presented a joint analysis of the constraints on the matter
power spectrum and the density of dark matter from three different
cosmological probes: the CMB temperature and polarization anisotropy
measurements of WMAP year three, the state-of-the-art weak lensing
COSMOS survey and two independent \lya forest data sets.  The
different observables are prone to very different systematic errors
and parameter degeneracies, and more importantly probe different
scales and redshifts.  Assessing their consistency is an important
test of the $\Lambda$CDM paradigm and is crucial for further improving
constraints on cosmological parameters. The measurements of the matter
power spectrum on small and intermediate scale based on \lya forest
and weak lensing data agree very well and suggest a higher amplitude
($\sig = 0.876 \pm 0.048$ with the analysis of SDSS \lya forest data
based on the flux derivatives method of \citet{Viel:2005ha}) than the
WMAP data alone ($\sig = 0.762\pm 0.046$).  The direction of
degeneracy between the amplitude of the power spectrum on galaxy
scales the parameters governing its shape (in other words, the
direction of degeneracy in $\sig$--$\ns$ and $\sig$--$\om$ space) is
different for the \lya forest and weak lensing data.  These two
observables thus complement each other very well and combining them
results in a significant improvement.  Combining all three observables
we get either $\sig=0.800\pm 0.023$, $\ns=0.971\pm 0.011$,
$\om=0.247\pm0.016$ (with the analysis of SDSS \lya forest data
based on the flux derivatives method of \citet{Viel:2005ha}) 
or $\sig=0.822\pm 0.032$, $\ns=0.960\pm 0.016$, $\om=0.282\pm0.026$
(with the high resolution \lya data of VHS).
We further explored the constraints for a running of the spectral index and
found the data  to be consistent with no running at less than
the 2$\sigma$ level. Adding the smaller scale data sets reduces the 
uncertainty on the running of the spectral index by a factor of two
with respect to WMAP alone.

\section*{ACKNOWLEDGMENTS}
Computations were done at the UK National Cosmology Supercomputer
Center funded by PPARC, HEFCE and Silicon Graphics / Cray Research and
at the HPCF (Cambridge High Performance Computer Cluster), as well as
on the MUST cluster in LAPP, Annecy (CNRS \& Universit\'e de Savoie).
JL acknowledges partial support from the EU 6th Framework Marie Curie
Research and Training network ``UniverseNet'' (MRTN-CT-2006-035863).

\vspace{\stretch{1}}


\newpage


\begin{thebibliography}{50}








\bibitem[\protect\citeauthoryear{Beltran, Garcia-Bellido, Lesgourgues \&
  Viel}{Beltran et~al.}{2005}]{Beltran:2005gr}
Beltran M.,  Garcia-Bellido J.,  Lesgourgues J.,    Viel M.,  2005, Phys. Rev.,
  D72, 103515, \eprint{astro-ph/0509209}

\bibitem[\protect\citeauthoryear{Croft et~al.,}{Croft
  et~al.}{2002}]{Croft:2000hs}
Croft R. A.~C.,  et~al., 2002, ApJ, 581, 20,
  \eprint{astro-ph/0012324}

\bibitem[\protect\citeauthoryear{Croft, Weinberg, Katz \& Hernquist}{Croft
  et~al.}{1998}]{Croft:1997jf}
Croft R. A.~C.,  Weinberg D.~H.,  Katz N.,    Hernquist L.,  1998, \apj, 495,
  44, \eprint{astro-ph/9708018}




\bibitem[\protect\citeauthoryear{Gnedin \& Hamilton}{Gnedin \&
  Hamilton}{2002}]{Gnedin:2001wg}
Gnedin N.~Y.,  Hamilton A. J.~S.,  2002, MNRAS, 334, 107,
  \eprint{astro-ph/0111194}

\bibitem[\protect\citeauthoryear{Gnedin \& Hui}{Gnedin \&
  Hui}{1998}]{Gnedin:1997td}
Gnedin N.~Y.,  Hui L.,  1998, MNRAS, 296, 44,
  \eprint{astro-ph/9706219}




\bibitem{step1}
Heymans C.\ et al., 2006, MNRAS 368, 1323

\bibitem{step2}
Massey R.\ et al., 2007, MNRAS 376, 13



\bibitem[\protect\citeauthoryear{Hinshaw et~al.,}{Hinshaw
  et~al.}{2006}]{Hinshaw:2006ia}
Hinshaw G.,  et~al., 2006, \eprint{astro-ph/0603451}

\bibitem[\protect\citeauthoryear{Hui et~al.,}{Hui  et~al.}{2001}]{Hui:2000rw}
Hui L.,  et~al., 2001, ApJ, 552, 15, \eprint{astro-ph/0005049}

\bibitem[\protect\citeauthoryear{Jena et~al.,}{Jena
  et~al.}{2005}]{Jena:2004fc}
Jena T.,  et~al., 2005, MNRAS, 361, 70,
  \eprint{astro-ph/0412557}

\bibitem[\protect\citeauthoryear{Kim, Viel, Haehnelt, Carswell \&
  Cristiani}{Kim et~al.}{2004}]{Kim:2003qt}
Kim T.~S.,  Viel M.,  Haehnelt M.~G.,  Carswell R.~F.,    Cristiani S.,  2004,
  MNRAS, 347, 355, \eprint{astro-ph/0308103}





\bibitem[\protect\citeauthoryear{McDonald}{McDonald}{2003}]{McDonald:2001fe}
McDonald P.,  2003, ApJ, 585, 34, \eprint{astro-ph/0108064}

\bibitem[\protect\citeauthoryear{McDonald et~al.,}{McDonald
  et~al.}{2000}]{McDonald:1999dt}
McDonald P.,  et~al., 2000, \apj, 543, 1, \eprint{astro-ph/9911196}


\bibitem[\protect\citeauthoryear{McDonald et~al.,}{McDonald
  et~al.}{2005}]{McDonald:2004xn}
McDonald P.,  et~al., 2005, ApJ, 635, 761, \eprint{astro-ph/0407377}

\bibitem[\protect\citeauthoryear{McDonald et~al.,}{McDonald
  et~al.}{2006}]{McDonald:2004eu}
McDonald P.,  et~al., 2006, ApJS, 163, 80, \eprint{astro-ph/0405013}

\bibitem[\protect\citeauthoryear{McDonald, Seljak, Cen, Bode \&
  Ostriker}{McDonald et~al.}{2005}]{McDonald:2004xp}
McDonald P.,  Seljak U.,  Cen R.,  Bode P.,    Ostriker J.~P.,  2005, Mon. Not.
  Roy. Astron. Soc., 360, 1471, \eprint{astro-ph/0407378}


\bibitem[\protect\citeauthoryear{Page et~al.,}{Page
  et~al.}{2006}]{Page:2006hz}
Page L.,  et~al., 2006, \eprint{astro-ph/0603450}






\bibitem[\protect\citeauthoryear{Spergel et~al.,}{Spergel
  et~al.}{2006}]{Spergel:2006hy}
Spergel D.~N.,  et~al., 2006, \eprint{astro-ph/0603449}



\bibitem[\protect\citeauthoryear{Springel}{Springel}{2005}]{Springel:2005mi}
Springel V.,  2005, MNRAS, 364, 1105,
  \eprint{astro-ph/0505010}

\bibitem[\protect\citeauthoryear{Springel, Yoshida \& White}{Springel
  et~al.}{2001}]{Springel:2000yr}
Springel V.,  Yoshida N.,    White S. D.~M.,  2001, New Astron., 6, 79,
  \eprint{astro-ph/0003162}


\bibitem[\protect\citeauthoryear{Viel \& Haehnelt}{Viel \&
  Haehnelt}{2006}]{Viel:2005ha}
Viel M.,  Haehnelt M.~G.,  2006, MNRAS, 365, 231,
  \eprint{astro-ph/0508177}

\bibitem[\protect\citeauthoryear{Viel, Haehnelt \& Springel}{Viel
  et~al.}{2004}]{Viel:2004bf}
Viel M.,  Haehnelt M.~G.,    Springel V.,  2004, MNRAS,
  354, 684, \eprint{astro-ph/0404600}


\bibitem[\protect\citeauthoryear{Viel, Haehnelt \& Springel}{Viel
  et~al.}{2006}]{Viel:2005eg}
Viel M.,  Haehnelt M.~G.,    Springel V.,  2006, MNRAS,
  367, 1655, \eprint{astro-ph/0504641}

\bibitem[\protect\citeauthoryear{Viel, Lesgourgues, Haehnelt, Matarrese \&
  Riotto}{Viel et~al.}{2005}]{Viel:2005qj}
Viel M.,  Lesgourgues J.,  Haehnelt M.~G.,  Matarrese S.,    Riotto A.,  2005,
  PRD, 71, 063534, \eprint{astro-ph/0501562}


\bibitem[\protect\citeauthoryear{Viel, Matarrese, Theuns, Munshi \& Wang}{Viel
  et~al.}{2003}]{Viel:2002gn}
Viel M.,  Matarrese S.,  Theuns T.,  Munshi D.,    Wang Y.,  2003, MNRAS,
  340, L47, \eprint{astro-ph/0212241}

\bibitem[\protect\citeauthoryear{Viel, Weller \& Haehnelt}{Viel
  et~al.}{2004}]{Viel:2004np}
Viel M.,  Weller J.,    Haehnelt M.,  2004, MNRAS, 355,
  L23, \eprint{astro-ph/0407294}




\bibitem[\protect\citeauthoryear{Zaldarriaga, Scoccimarro \& Hui}{Zaldarriaga
  et~al.}{2003}]{Zaldarriaga:2001xs}
Zaldarriaga M.,  Scoccimarro R.,    Hui L.,  2003, ApJ, 590, 1,
  \eprint{astro-ph/0111230}







\bibitem{lewis} Lewis A., Bridle S.,  2002, Phys. Rev., D66, 103511; http://www.cosmologist.info


\bibitem{seljak}
Seljak U., Slosar A., McDonald P., 2006, JCAP, 0610, 014, [arXiv: astro-ph/0604335] 




\bibitem{seljakmakarov}
Seljak U., Makarov A., McDonald P., Trac H., 2006, PhysRevLett, 97, 191303, [arXiv: astro-ph/0602430] 


\bibitem{vielsterile}
 Viel M.,  Lesgourgues J.,  Haehnelt M.~G.,  Matarrese S.,    Riotto A.,  2006, PhysRevLett, 97, 071301, [arXiv: astro-ph/0605706] 




\bibitem{vielhaehnelt06} Viel M., Haehnelt M.G., 2006, MNRAS, 365, 231, 
[arXiv:astro-ph/0508177]


\bibitem{heavens}
Heavens A.,  Kitching T., Taylor A., 2006, [arXiv: astro-ph/0606568]

\bibitem[\protect\citeauthoryear{Leauthaud et al.}{2007}]{apjse_lea}
Leauthaud A.\ et al., 2007, [arXiv: astro-ph/0702359]

\bibitem{massey}
Massey R. et al., 2007, [arXiv: astro-ph/0701480], accepted in ApJ

\bibitem{massey2}
Massey R. et al., 2007, Nature 445, 286

\bibitem[\protect\citeauthoryear{Mobasher et al.}{2007}]{apjse_mob}
Mobasher B.\ et al., 2007, ApJS in press

\bibitem[\protect\citeauthoryear{Semboloni et al.}{2006}]{semvariance}
Semboloni E., van Waerbeke L., Heymans C., Hamana T., Colombi S., 
White M.\ \& Mellier Y., 2006, MNRAS, submitted (astro-ph/0606648)

\bibitem{smith}
Smith R.E., et al., 2003, MNRAS, 341, 1131, [arXiv: astro-ph/0207664)]

\bibitem{Lewis:1999bs}
  A.~Lewis, A.~Challinor and A.~Lasenby,
  Astrophys.\ J.\  {\bf 538} (2000) 473
  [arXiv:astro-ph/9911177].

\bibitem{Rudd:2007zx}
  D.~H.~Rudd, A.~R.~Zentner and A.~V.~Kravtsov,
  arXiv:astro-ph/0703741.

\bibitem{viel06}
Viel M., Haehnelt M.G., Lewis, 2006, MNRAS, 370, 51L, [arXiv: astro-ph/0604310]

\bibitem{cosmosall}
Scoville N.,  et al., 2007a, [arXiv:astro-ph/0612305]; Scoville N.,  et al., 2007b, [arXiv:astro-ph/0612306]; Massey R. et al., 2007, Nature, 445, 286


\bibitem{benjamin}
Benjamin J., Heymans C., Semboloni E., Van Waerbeke L., Hoekstra H., Erben T., Gladders M.D., Hetterscheidt M., Mellier Y., Yee H.K.C., 2007, [arXiv:astro-ph/0703570]

\bibitem{bergeron}
Bergeron J., et al., 2004, The large programme "cosmic evolution of the IGM", 
The Messenger, 118, 40 

\bibitem{BBKS}
  J.~M.~Bardeen, J.~R.~Bond, N.~Kaiser and A.~S.~Szalay,
  Astrophys.\ J.\  {\bf 304} (1986) 15.



\end{thebibliography}
\end{document}